\documentclass[twocolumn,twoside,slac_two]{revtex4}
\usepackage{graphicx}
\usepackage{fancyhdr}
\def\mincir{\raise -2.truept\hbox{\rlap{\hbox{$\sim$}}\raise5.truept \hbox{$<$}\ }}
\def\mincireq{\hbox{\raise0.5ex\hbox{$<\lower1.06ex\hbox{$\kern-1.07em{\sim}$}$}}}
\def\magcir{\raise-2.truept\hbox{\rlap{\hbox{$\sim$}}\raise5.truept \hbox{$>$}\ }}
\def\gr{\kern 2pt\hbox{}^\circ{\kern -2pt K}} % ====> GRADI KELVIN
\begin{document}

\title{Blazars as beamlights to probe the Extragalactic Background Light \\
	 in the Fermi and Cherenkov telescopes era}

\author{M.\,Persic} 
\affiliation{INAF-Trieste and INFN-Trieste, via G.B.\,Tiepolo 11, I-34143 Trieste TS, Italy}
\author{N.\,Mankuzhiyil} 
\affiliation{Udine University and INFN-Trieste, via delle Scienze 208, I-33100 Udine UD, Italy}
\author{F.\,Tavecchio} 
\affiliation{INAF-Brera, via E.\,Bianchi 46, I-23807 Merate LC, Italy}

\begin{abstract}
The Extragalactic Background Light (EBL) is the integrated light from all the 
stars that have ever formed, and spans the IR-UV range. The interaction of 
very-high-energy (VHE: $E>100\,$GeV) $\gamma$-rays, emitted by sources located 
at cosmological distances, with the intervening EBL results in $e^-e^+$ pair 
production that leads to energy-dependent attenuation of the observed VHE flux. 
This introduces a fundamental ambiguity in the interpretation of the measured 
VHE blazar spectra: neither the intrinsic spectra, nor the EBL, are separately 
known -- only their combination is. 
In this paper we propose a method to measure the EBL photon number density. 
It relies on using simultaneous observations of blazars in the optical, X-ray, 
high-energy (HE: $E>100\,$MeV) $\gamma$-ray (from the Fermi telescope), and 
VHE\,$\gamma$-ray (from Cherenkov telescopes) bands. For each source, the method 
involves best-fitting the spectral energy distribution (SED) from optical through 
HE\,$\gamma$-rays (the latter being largely unaffected by EBL attenuation as long 
as $z \mincir 1$) with a Synchrotron Self-Compton (SSC) model. We extrapolate 
such best-fitting models into the VHE regime, and assume they represent the blazars' 
intrinsic emission. Contrasting measured versus intrinsic emission leads to a 
determination of the $\gamma$-$\gamma$ opacity to VHE photons -- hence, upon 
assuming a specific cosmology, we derive the EBL photon number density. Using, for 
each given source, different states of emission will only improve the accuracy of 
the proposed method. We demonstrate this method using recent simultaneous 
multi-frequency observations of the blazar PKS\,2155-304 and discuss how similar 
observations can more accurately probe the EBL.  
\end{abstract}

\maketitle

\thispagestyle{fancy}

\section{INTRODUCTION}

The Extragalactic Background Light (EBL), in both its level and degree of cosmic 
evolution, reflects the time integrated history of light production and re-processing 
in the Universe, hence the history of cosmological star-formation. Roughly speaking, 
its shape must reflect the two humps that characterize the spectral energy distributions 
(SEDs) of galaxies: one arising from starlight and peaking at $\lambda \sim 1\,\mu$m 
(optical background), and one arising from warm dust emission and peaking at $\lambda 
\sim 100\,\mu$m (infrared background). 

Direct measurements of the EBL are hampered by the dominance of foreground 
emission (interplanetary dust and Galactic emission), hence the level of EBL 
emission is uncertain by a factor of several. 

One approach has been modeling the EBL arising from an evolving population of 
galactic stellar populations: however, uncertainties in the assumed galaxy 
formation and evolution scenarios, stellar initial mass function, and star 
formation rate have led to significant discrepancy among models (e.g., 
\cite{SalSte98, SteJag98, Kne+02, Kne+04, Ste+06}). These models have been 
used to correct observed VHE spectra and deduce (EBL model dependent) 
'intrinsic' VHE\,$\gamma$-ray emissions. 

The opposite approach, of a more phenomenological kind, deduces upper limits 
on the level of EBL attenuation making basic assumptions on the intrinsic 
VHE\,$\gamma$-ray shape of AGN spectra: assuming, specifically, that the VHE 
photon index must be $\Gamma \geq 1.5$; e.g., (\cite{Ahar+06, MazGoeb07, 
MazRau07}); but see (\cite{Ste+07}), or that the same-slope extrapolation of 
the observed {\it Fermi}/LAT HE spectrum into the VHE domain exceeds the intrinsic 
VHE spectrum there (\cite{Georg+09}). The only unquestionable constraints on 
the EBL are model-independent lower limits based on galaxy counts (\cite{Dole+06, 
Fran+08}). It should be noted, however, that the EBL upper limits in the 
2--80$\mu$m obtained by \cite{MazRau07} combining results from all known TeV 
blazar spectra (based on the assumption that the intrinsic $\Gamma \geq 1.5$) 
are only a factor $\approx$2--2.5 above the absolute lower limits from source 
counts. So it would appear that there is little room for additional components 
like Pop\,III stars, unless we miss some fundamental aspects of blazar emission 
theory (which we never observed in local sources, however).

An attempt to measure the EBL used the relatively faraway blazar 3C\,279 as a 
background light source (\cite{Ste+92}), assumed that the intrinsic VHE spectrum 
was known from modeling and extrapolating the (historical) average broad-band 
data. However, blazars are highly variable sources, so it's almost impossible to 
determine with confidence the intrinsic TeV spectrum -- which itself can be variable. 

In this paper we propose a method to measure the EBL that improves on \cite{Ste+92} 
by making a more realistic assumption on the intrinsic TeV spectrum. Simultaneous 
optical/X-ray/HE/VHE (i.e., eV/keV/GeV/TeV) data are crucial to this method, 
considering the strong and rapid variability displayed by most blazars. After 
reviewing features of EBL absorption (sect.\,2) and of the SSC emission model 
(sect.\,3), in sect.\,4 we describe our technique, in sect.\,5 we apply it to 
recent multifrequency observations of PKS\,2155-304 and determine the photon-photon 
optical depth out to that source's redshift. In sect.\,6 we discuss our results.

\section{EBL ABSORPTION}

The cross section for the reaction $\gamma \gamma$$\rightarrow$$e^{\pm}$ is (\cite{Heit60}),
\begin{eqnarray}
\lefteqn{
\sigma_{\gamma \gamma}(E, \epsilon) ~=~ {3 \over 16}\sigma_{\rm T} ~(1-\beta^2)~~\times }
	\nonumber\\
 & & 
\times ~~ \biggl[\, 2\,\beta\,(\beta^2-2) ~+~ 
(3-\beta^4) ~{\rm ln} {1+\beta \over 1-\beta} \, \biggr]\, , 
\end{eqnarray}
where $\sigma_{\rm T}$ is the Thompson cross section and 
$\beta$$\equiv$$\sqrt{1-(m_ec^2)^2/E\epsilon}$. For demonstration purposes let 
us assume, following \cite{Ste+92}, that $n(\epsilon)$ is the local number density of 
EBL photons having energy equal to $\epsilon$ (no redshift evolution -- as befits the 
relatively low redshifts accessible to IACTs), $z_e$ is the source redshift, and 
$\Omega_0$$=$1: the corresponding optical depth due to pair creation attenuation 
between the source and the Earth, is (see (\cite{Ste+92}) 
\begin{eqnarray}
\lefteqn{
\tau_{\gamma \gamma}(E, z_e) ~=~ {c \over H_0} \int_0^{z_e} \sqrt{1+z}~ {\rm d}z ~ 
\int_0^2 {x \over 2} {\rm d}x ~~~ \times} 
                \nonumber\\
 & & 
\times ~ \int_{2(m_ec^2)^2 \over Ex(1+z)^2}^\infty n(\epsilon)~~ \sigma_{\gamma 
\gamma}\bigl(2xE \epsilon (1+z)^2\bigr) ~~ {\rm d}\epsilon\,,
\end{eqnarray}
where $x$$\equiv$$(1$$-$${\rm cos}\,\theta)$, $\theta$ being the angle between the photons, 
and $H_0$ is the Hubble constant. We further assume, again following \cite{Ste+92}, that the 
local EBL spectrum has a power-law form, $n(\epsilon)$$\propto$$\epsilon^{-2.55}$. Then 
Eq.(1) yields $\tau(E,z)$$\propto$$E^{1.55}$$z_s^{\eta}$ with $\eta$$\sim$1.5. 

This calculation, although it refers to an idealized and somewhat simplified situation, 
highlights an important property of the VHE flux attenuation by the $\gamma_{\rm VHE}$$
\gamma_{\rm EBL}$$\rightarrow$e$^+$e$^-$ interaction: $\tau_{\gamma\gamma}$ depends both 
on the distance traveled by the VHE photon (hence on $z$) and on its (measured) energy $E$.
So the spectrum measured at Earth is distorted with respect to the emitted spectrum.
In detail, the expected VHE $\gamma$-ray flux at Earth will be: $F(E)$$=($d$I/$d$E)\,e^
{-\tau_{\gamma \gamma}(E)}$ (differential) and $F($$>$$E)$$=$$\int_E^\infty ($d$I/$d$E^
\prime)\,e^{-\tau_{\gamma \gamma}(E^{\prime})}dE^\prime$ (integral).

\section{BLAZAR SSC EMISSION}

In order to reduce the degrees of freedom, we use a simple one-zone SSC model 
(for details see \cite{Tav+98, TavMar03}). This has been shown to adequately 
describe broad-band SEDs of most blazars (\cite{Tav+98, Ghis+98}) and, for a 
given blazars, both its ground and excited states (\cite{Tagl+08}). The reason 
for the one-zone model to work is that in most blazars the temporal variability 
is clearly dominated by one characteristic timescale, which implies one dominant 
characteristic size of the emitting region (\cite{Ander+09}).

The emission zone is supposed to be spherical with radius $R$, in motion with bulk 
Lorentz factor $\Gamma$ at an angle $\theta $ with respect to the line of sight. 
Special relativistic effects are described by the relativistic Doppler factor, 
$\delta=[\Gamma(1-\beta\,{\rm cos}\,\theta)]^{-1}$. The energy spectrum of the 
emitting relativistic electrons is described by a smoothed broken power-law 
function of the electron Lorentz factor $\gamma$, with limits $\gamma _1$ and 
$\gamma _2$ and break at $\gamma _{\rm br}$. In calculating the SSC emission we 
use the full Klein-Nishina cross section. 

As detailed in \cite{Tav+98}, this simple model can be fully constrained by using 
simultaneous multifrequency observations. Indeed, the total number of free 
parameter of the model is reduced to 9: the 6 parameter specifying the electron 
energy distribution, plus the Doppler factor $\delta$, the size of the emission 
region $R$, and the magnetic field $B$. On the other hand, from observations 
ideally one can derive 9 observational quantities: the slopes of the synchrotron 
bump after and above the peak $\alpha _{1,2}$ (uniquely connected to $n_{1,2}$), 
the synchrotron and SSC peak frequencies ($\nu _{\rm s,C}$) and luminosities 
$L_{\rm s,C}$, and the minimum variability timescale $t_{\rm var}$ which provides 
an estimate of the size of the sources through $R < c t_{\rm var} \delta $. 

Therefore, once the relevant observational quantities are known, one can uniquely 
derive the set of SSC parameters.

\section{THE METHOD}

The method we are proposing stems from the consideration that both the 
EBL and the intrinsic VHE\,$\gamma$-ray spectra of background sources are 
fundamentally unknown. In order to measure the EBL at different $z$, one should single out 
a class of sources that is homogeneous, i.e. it can be described by one 
same emission model at all redshifts. This approach is meant to minimize biases that may 
possibly arise from systematically different SED modelings adopted for 
different classes of sources at different distances. So we choose the class of source that both 
has the simplest emission model and has the potentiality of being seen 
from large distances: blazars, i.e. the AGN whose relativistic jets  
point toward the observer so their luminosities are boosted by a large 
factor and dominate the source flux with their SSC emission. Within 
blazars, we propose to use the sub-class of "high-frequency peaked BL 
Lacs" (HBL), because their Compton peak can be more readily detected by 
IACTs than other types of blazar, and because their HE spectrum can be 
described as a single (unbroken) power law in photon energy, unlikely 
other types of blazar (\cite{Lott09, Abdo+09}). 

For a given blazar, our method relies on using, a simultaneous broad-band 
SED that samples the optical, X-ray, high-energy (HE: $E>100\,$MeV) 
$\gamma$-ray (from the {\it Fermi} telescope), and VHE\,$\gamma$-ray 
(from Cherenkov telescopes) bands. A given SED will be best-fitted, 
from optical through HE\,$\gamma$-rays, with a Synchrotron Self-Compton 
(SSC) model. [Photons with $E \mincir 100$\,GeV are largely unaffected 
by EBL attenuation (for reasonable EBL models) as long as $z \mincir 1$.] 
Extrapolating such best-fitting 
SED model into the VHE regime, we shall assume it represents the blazar's 
intrinsic emission. Contrasting measured versus intrinsic emission yields 
a determination of $e^{-\tau_{\gamma\gamma}(E,\,z)}$, the energy-dependent 
absorption of the VHE emission coming from a source located at redshift $z$ 
due to pair production with intervening EBL photons. Upon assumption of a 
specific cosmology, the final step is deriving the EBL photon number density. 

Using, for each blazar, SEDs from different states of emission will improve 
the accuracy of the method by increasing the number of EBL measurements at 
that redshift. 

\subsection{Best-fit procedure: $\chi^2$ minimization}

In order to fit the observed optical, X-ray and HE $\gamma$-ray flux with 
the SSC model, a $\chi^{2}$ minimization is used. We vary the SSC model's 
9 parameters by small logarithmic steps. If the variability timescale of 
the flux, $t_{\rm var}$, is known, one can set $R \sim c t_{\rm var} \delta$, 
so the free parameters are reduced to 8.
We assume here $\gamma_{\rm min}$$=$1: for a plasma with $n_{\rm e}$$\approx${\cal 
O}($10$)\,cm$^{-3}$ and $B$$\approx${\cal O}(0.1)\,G (as generally appropriate for 
TeV blazar jets: e.g., \cite{Ghis+98, CostGhis02, Fink+08}), this approximately 
corresponds to the energy below which Coulomb losses exceed the synchrotron losses 
(e.g., \cite{Reph79, Sar99}) and hence the electron spectrum bends over and no 
longer is power-law. However, in general $\gamma_{\rm min}$ should be left to vary -- 
e.g., cases of a "narrow" Compton component require $\gamma_{\rm min}$$>$1 (\cite{Tav+09}). 
In order to reduce the run time of the code, the steps are adjusted in each run 
such that, a larger $\chi^2$ is followed by larger steps. 

% §§§§§§§§§§§§§§§§§§§§§§§§§§§§ FIGURE 1 §§§§§§§§§§§§§§§§§§§§§§§§§§§§§§§§§§
\begin{figure*}[t]
\centering
\includegraphics[width=125mm]{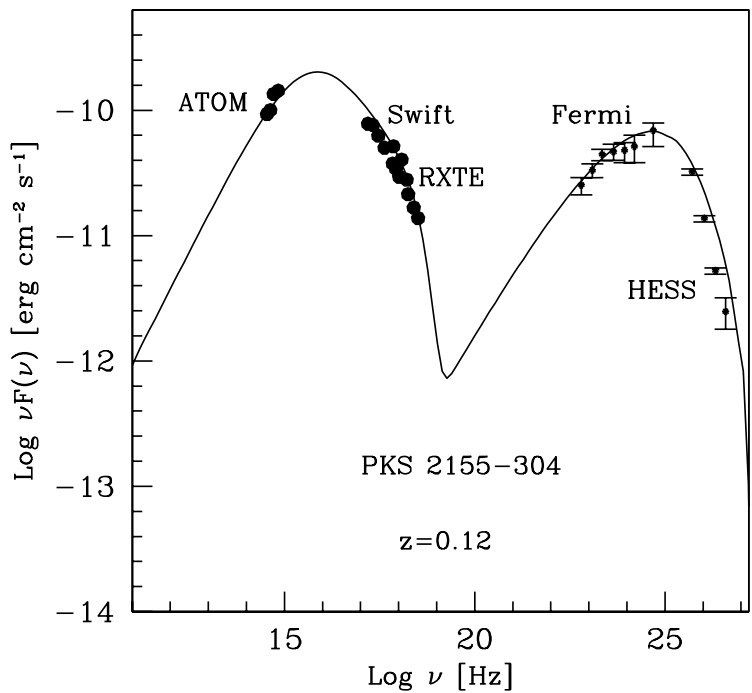}
\caption{
Data (symbols: from \citep[see][]{Ahar+09}) and 
best-fit SSC model (solid curve) of the SED of PKS\,2155-304. 
The best-fit SSC parameters are:  
$n_{\rm e}=150$\,cm$^{-3}$, 
$\gamma_{\rm br}=2.9 \times 10^4$,  
$\gamma_{\rm max}=8 \times 10^5$,
$\alpha_1=1.8$,
$\alpha_2=3.8$,
$R=3.87 \times 10^{16}$\,cm, 
$\delta=29.2$, 
$B=0.056$\,G.
The obtained values of $R$ and $\delta$ imply a variability 
timescale $t_{\rm var} \sim R /(c \delta)$, which is 
compatible with the observed value of $\approx$12\,hr. 
} 
\label{sed}
\end{figure*}
% §§§§§§§§§§§§§§§§§§§§§§§§§§§§§§§§§§§§§§§§§§§§§§§§§§§§§§§§§§§§§§§§§§§§§§§§

\section{RESULTS: APPLICATION TO PKS\,2155-304}

We apply the procedure described in Sect.\,4 to the simultaneous SED data set 
of PKS\,2155-304 described in \cite{Ahar+09}. The data and resulting best-fit 
SSC model (from optical through HE\,$\gamma$-rays) are shown in Fig.(\ref{sed}). 
The extrapolation of the model into the VHE\,$\gamma$-ray range clearly lies below 
the observational H.E.S.S. data, progressively so with increasing energy. We 
attribute this effect to EBL attenuation, $F_{\rm obs}(E;\,z)$$=$$F_{\rm 
em}(E;\,z)\, e^{-\tau_{\gamma\gamma}(E;\,z)}$. The corresponding values of 
$\tau_{\gamma\gamma}(E;\,z)$ for $E$$=$0.23, 0.44, 0.88, 1.70\,TeV and 
$z$$=$0.12 are, respectively, $\tau_{\gamma\gamma}=0.12$, 0.48, 0.80, and 0.87 . 

We note that the SED analysis of \cite{Ahar+09} was based on a slightly different 
SSC model, that involved a three-slope (as opposed to our two-slope) electron 
spectrum. This difference may lead to a somewhat different decreasing wing of the 
modeled Compton hump, and hence to a systematic difference in the derived $\tau_{
\gamma\gamma}(E;\,z)$. That said, it's however interesting to note that the main 
parameters describing the plasma blob ($B$, $\delta$, $n_{\rm e}$) take on similar 
values in our best-fit analysis and in \cite{Ahar+09}. 

\begin{figure*}[t]
\centering
\includegraphics[width=80mm]{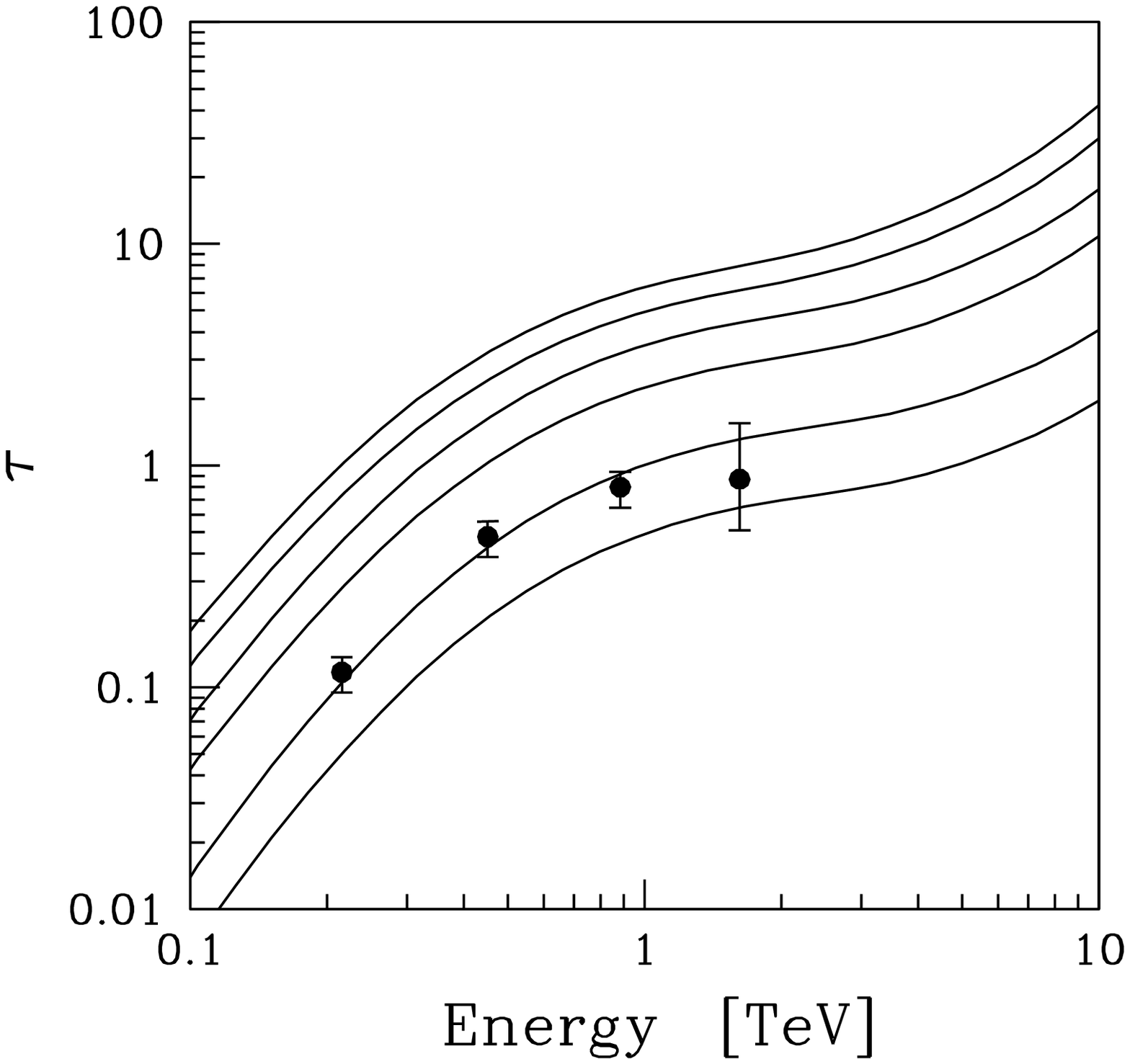}
\includegraphics[width=80mm]{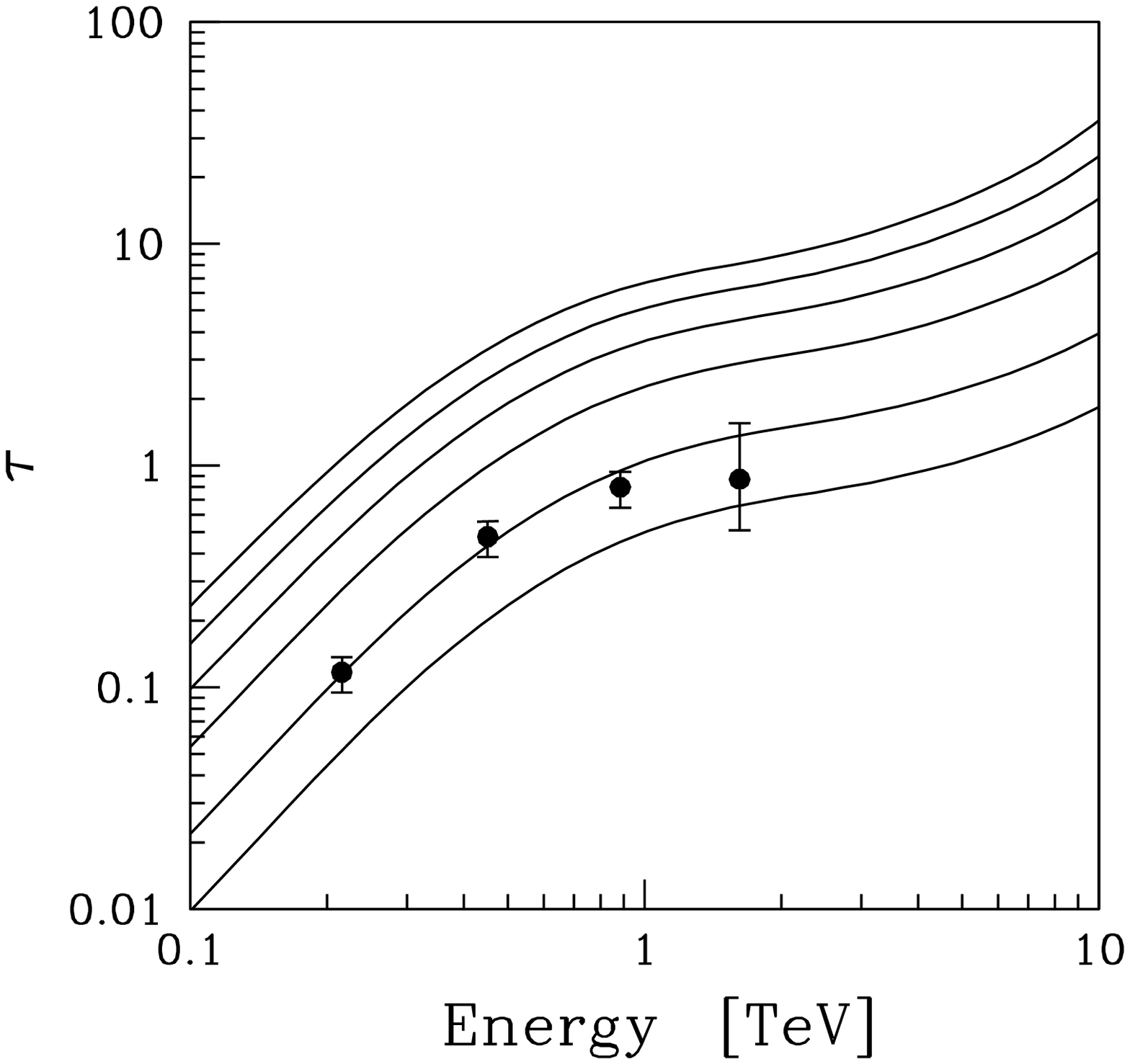}
\includegraphics[width=80mm]{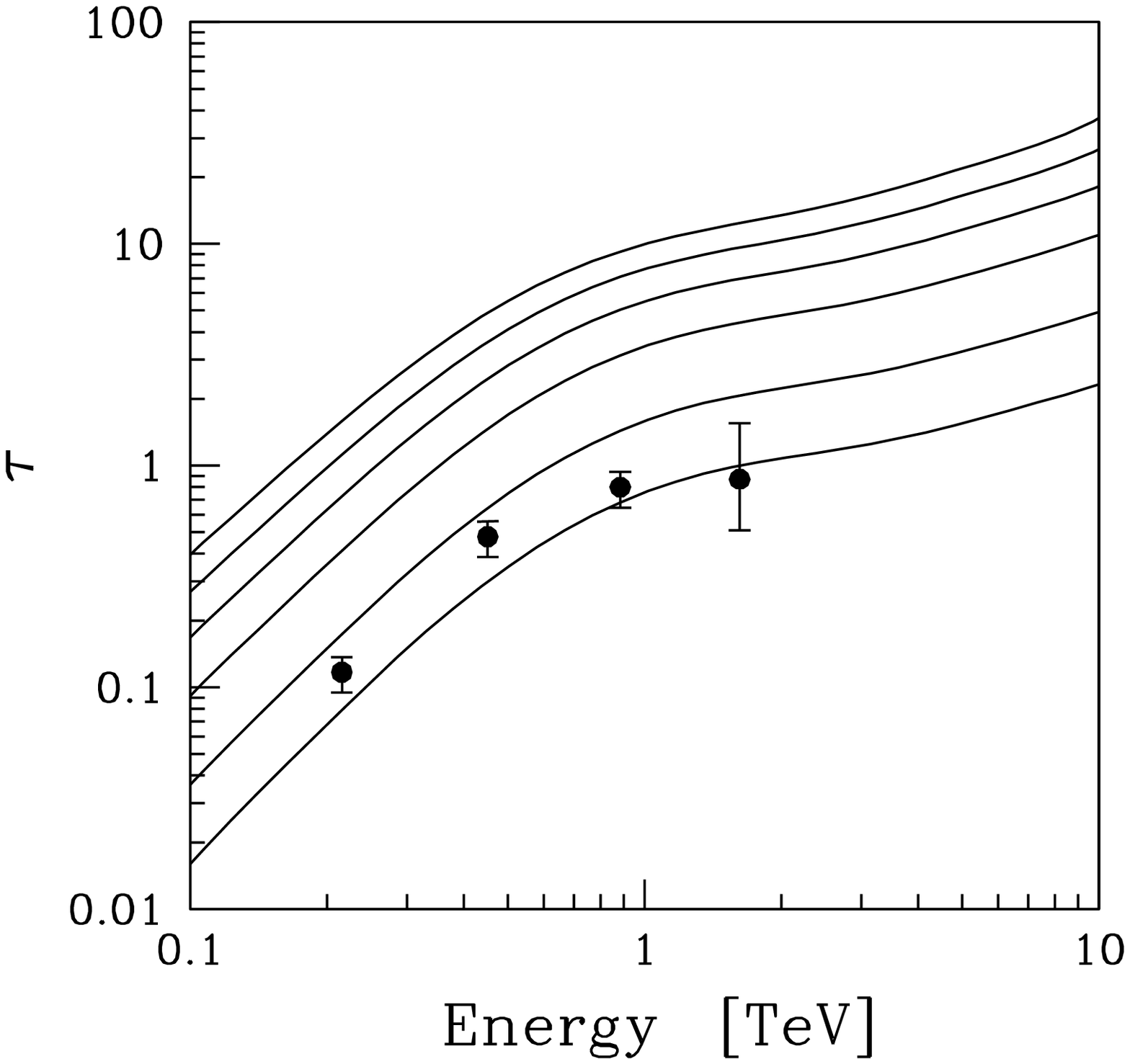}
\includegraphics[width=80mm]{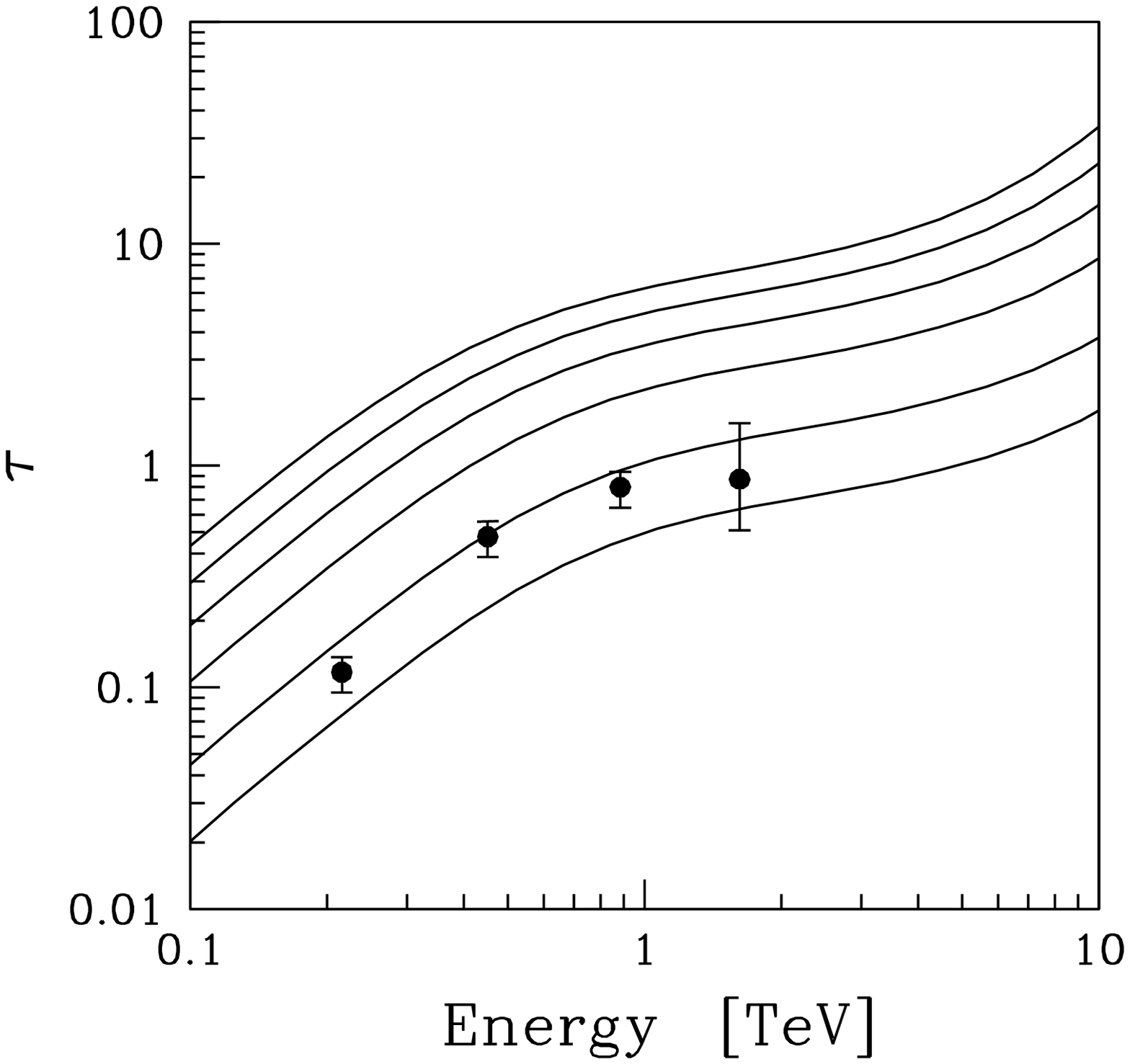}
\caption{
Measured values of $\tau_{\gamma\gamma}(E;\,z)$ for $E$$=$0.23, 0.44, 0.88, 
1.70\,TeV derived from comparing, for simultaneous observations of the HBL 
blazar PKS\,2155-304 ($z$$=$0.12), the (EBL-affected) VHE\,$\gamma$-ray data 
with the eV-through-GeV best-fitting SSC model extrapolated into the TeV domain. 
The curves represent, for redshifts $z=0.05$, 0.1, 0.2, 0.3, 0.4, and 0.5, the 
optical depth $\tau_{\gamma\gamma}(E)$ according to 
\cite{Fran+08} ({\it top left}), and upper limits to it according to 
\cite{Gilm+09} ({\it top right}), 
\cite{Kne+04} ({\it bottom left}), 
\cite{RauMaz08} ({\it bottom right}). 
} 
\label{tau}
\end{figure*}

In Fig.(\ref{tau}) we compare our determination of $\tau_{\gamma\gamma}$ with some 
recent results (\cite{Fran+08}) or upper limits (\cite{Kne+04, RauMaz08, Gilm+09}). 
Whereas our values are generally compatible with previously published constraints, 
we note that our values closely agree with the corresponding values of \cite{Fran+08}, 
which are derived from galaxy number counts and hence represent the light contributed 
by the stellar populations of galaxies prior to the epoch corresponding to source redshift 
$z_{\rm s}$ -- i.e., the minimum amount (i.e., the guaranteed level) of EBL. 

% §§§§§§§§§§§§§§§§§§§§§§§§§§§§ FIGURE 2 §§§§§§§§§§§§§§§§§§§§§§§§§§§§§§§§§§
% §§§§§§§§§§§§§§§§§§§§§§§§§§§§§§§§§§§§§§§§§§§§§§§§§§§§§§§§§§§§§§§§§§§§§§§§

\section{DISCUSSION} 

The method for measuring the EBL we have proposed in this paper is admittedly 
model-dependent. However, its only requirement is that all the sources used as 
background beamlights should have one same emission model. In the application 
proposed here, we have used a one-zone SSC model where the electron spectrum was 
a (smoothed) double power law applied to the SED of the HBL blazar PKS\,2155-304. 
While this choice was encouraged by the current observational evidence fact that 
seem to HBLs have, with no exception, single-slope {\it Fermi}-LAT spectra 
(\cite{Lott09}), we could have as well adopted the choice (\cite{Ahar+09}) of a 
triple power law electron spectrum in our search for the best-fit SSC model of 
PKS\,2155-304's SED. Should the latter electron distribution be shown to provide 
a better fit to HBL {\it Fermi}-LAT spectra, then it would become our choice. In 
general, what matters to the application of this method, is that {\it all} source 
SEDs be fit with one same SSC model.

This work will be the subject of a forthcoming paper (\cite{Mank+10}).

\end{document}